\documentclass[twocolumn,pre,superscriptaddress]{revtex4}

\usepackage{graphicx}
\usepackage{amsmath,amssymb}
\usepackage{color}
\usepackage{calrsfs}

\newcommand\defn{\textit}
\newcommand\mat{\mathbf}

\newcommand\dd{\mathrm{d}}

\renewcommand\vec{\mathbf}
\newcommand\av[1]{\langle#1\rangle}
\newcommand\etal{\textit{et~al.}}
\newcommand\half{\tfrac12}

\renewcommand{\Im}{\operatorname{Im}}

\newcommand{\tr}{\operatorname{Tr}}

\newcommand{\ccdot}{\!\cdot\!}

\newcommand\pin{p_\textrm{in}}
\newcommand\pout{p_\textrm{out}}

\begin{document}

\title{Spectra of random graphs with community structure and arbitrary
  degrees}

\author{Xiao Zhang}
\affiliation{Department of Physics, University of Michigan, Ann Arbor, MI
  48109}
\author{Raj Rao Nadakuditi}
\affiliation{Department of Electrical Engineering and Computer Science,
  University of Michigan, Ann Arbor, MI 48109}
\author{M. E. J. Newman}
\affiliation{Department of Physics, University of Michigan, Ann Arbor, MI
  48109}
\affiliation{Center for the Study of Complex Systems, University of
  Michigan, Ann Arbor, MI 48109}

\begin{abstract}
  Using methods from random matrix theory researchers have recently
  calculated the full spectra of random networks with arbitrary degrees and
  with community structure.  Both reveal interesting spectral features,
  including deviations from the Wigner semicircle distribution and phase
  transitions in the spectra of community structured networks.  In this
  paper we generalize both calculations, giving a prescription for
  calculating the spectrum of a network with both community structure and
  an arbitrary degree distribution.  In general the spectrum has two parts,
  a continuous spectral band, which can depart strongly from the classic
  semicircle form, and a set of outlying eigenvalues that indicate the
  presence of communities.
\end{abstract}


\maketitle

\section{Introduction}
\label{sec:intro}
Spectral analysis of networks provides a useful complement to traditional
analyses that focus on local network properties like degree distributions,
correlation functions, or subgraph densities~\cite{Newman03d,Boccaletti06}.
Spectral analysis can return nonlocal information about network structure
such as optimal partitions~\cite{Fiedler73,PSL90}, community
structure~\cite{Newman06c}, and nonlocal centrality
measures~\cite{Bonacich87} and has been widely used in the study of
real-world network data since the 1970s.  In additional to the development
of practical algorithms and methods based on network spectra, such as
spectral partitioning schemes and community detection algorithms, a
considerable amount of work has been done on the analytic calculation of
spectra for synthetic networks generated using random
models~\cite{FDBV01,GKK01a,CLV03,DGMS03,Kuhn08,CGO09,NN12,NN13}.  Study of
these model networks can help us to understand how particular features of
network structure are reflected in spectra and to anticipate the
performance of spectral algorithms.

Recent work on the spectra of networks with community structure, for
instance, has demonstrated the presence of a ``detectability threshold'' as
a function of the strength of the embedded structure~\cite{NN12}.  When the
community structure becomes sufficiently weak it can be shown that the
spectrum loses all trace of that structure, implying that any method or
algorithm for community detection based on spectral properties must fail at
this transition point.  A limitation of this work, however, is that the
synthetic networks studied have Poisson degree distributions, which makes
the calculations easier but is known to be highly unrealistic; real-world
degree distributions are very far from Poissonian.

In other work a number of authors have studied the spectra of synthetic
networks having broad degree distributions, such as the power-law
distributions observed in many real-world
networks~\cite{FDBV01,GKK01a,CLV03,DGMS03,Kuhn08,NN13}.  Among other
results, it is found that while the spectrum for Poisson degree
distributions follows the classic Wigner semicircle law, in the more
general case it departs from the semicircle, sometimes dramatically.

In this paper, we combine these two previous lines of investigation and
study the spectra of networks that possess general degree distributions and
simultaneously contain community structure.  To do this, we make use of a
recently proposed network model that generalizes the models studied before.
We derive an analytic prescription for calculating the adjacency matrix
spectra of networks generated by this model, which is exact in the limit of
large network size and large average degree.  In general the spectra have
two components.  The first is a continuous spectral band containing most of
the eigenvalues but having a shape that deviates from the semicircle law
seen in networks with Poisson degree distribution.  The second component
consists of outlying eigenvalues, outside the spectral band and normally
equal in number to the number of communities in the network.

\section{The model}
The previous calculations described in the introduction make use of two
classes of model networks.  For networks with community structure,
calculations were performed using the \defn{stochastic block model}, in
which vertices are divided into groups and edges placed between them
independently at random with probabilities that depend on the group
membership of the vertices
involved~\cite{HLL83,CK01,RL08,DKMZ11a,NN12,HRN12}.  This model gives
community structure of tunable strength but vertices have a Poisson
distribution of degrees within each community.

For networks without community structure but with non-Poisson degree
distributions, most calculations have been performed using the so-called
configuration model, a random graph conditioned on the actual degrees of
the vertices~\cite{MR95,NSW01}, or a variant of the configuration model in
which one fixes only the expected values of the degrees and not their
actual values~\cite{CL02b}.

The calculations presented in this paper make use of a model proposed by
Ball~\etal~\cite{BKN11} that simultaneously generalizes both the stochastic
block model and the configuration model, so that both are special cases of
the more general model.  The model of Ball~\etal\ is defined as follows.
We assume an undirected network of~$n$ vertices labeled $i=1\ldots n$, with
each of which is associated a $q$-component real vector~$\vec{k}_i$
where~$q$ is a parameter we choose.  Then the number of edges between
vertices~$i$ and~$j$ is an independent, Poisson-distributed random variable
with mean $\vec{k}_i\ccdot\vec{k}_j/2m$, where $m$ is a normalizing
constant given by
\begin{equation}
2m = \Biggl|\,\sum_{i=1}^n \vec{k}_i \Biggr|.
\label{eq:twom}
\end{equation}
Physically the value of $m$ represents the average total number of edges in
the whole network.  Its inclusion is merely conventional---one could easily
omit it and renormalize~$\vec{k}_i$ accordingly, and in fact Ball~\etal\
did omit it in their original formulation of the model.  However, including
it will simplify our notation later, as well as making the connection
between this model and the configuration model clearer.

The expected number of edges between vertices must be non-negative and
Ball~\etal\ ensured this by requiring that the elements of the
vectors~$\vec{k}_i$ all be non-negative, but this is not strictly necessary
since one can always rotate the vectors globally through any angle (thereby
potentially introducing some negative elements) without affecting their
products~$\vec{k}_i\ccdot\vec{k}_j$.  In this paper we will only require
that all products be nonnegative, which includes all cases studied by
Ball~\etal\ but also allows us to consider some cases they did not.

Note that it is possible in this model for there to be more than one edge
between any pair of vertices (because the number of edges is Poisson
distributed) and this may seem unrealistic, but in almost all real-world
situations we are concerned with networks that are sparse, in the sense
that only a vanishing fraction of all possible edges is present in the
network, which means that $\vec{k}_i\cdot\vec{k}_j/2m$ will be vanishing as
$n$ becomes large.  We will assume this to be the case here, in which case
the chances of having two or more edges between the same pair of vertices
also vanishes and for practical purposes the network contains only single
edges.

The average degree~$c$ of a vertex in the network is
\begin{equation}
c = {2m\over n} = \biggl| {1\over n} \sum_{i=1}^n \vec{k}_i \biggr|,
\end{equation}
and hence increases in proportion to the average of~$\vec{k}_i$.  In this
paper we will consider networks where the vectors~$\vec{k}_i$ can have a
completely general distribution, which gives us a good deal of flexibility
about the structure of our network, but consider for example a network in
which the vectors have arbitrary lengths, but each one points toward one of
the corners of a regular $q$-simplex in a (hyper)plane perpendicular to
the direction~$(1,1,1,\ldots)$.  For such a choice the vectors have the
form $\vec{k}_i = k_i \vec{v}_r$, where $k_i$ is the magnitude of the
vector and $\vec{v}_r$ is one of $q$ unit vectors that will denote the
group~$r$ that vertex~$i$ belongs to.  Then
\begin{equation}
\vec{k}_i\ccdot\vec{k}_j = k_ik_j \vec{v}_r\ccdot\vec{v}_s
  = k_ik_j [\delta_{rs} + (1-\delta_{rs})\cos\phi],
\end{equation}
where $\phi$ is the angle between unit vectors~$\vec{v}_r$ and~$\vec{v}_s$
(all vectors being separated by the same angle in a regular simplex).  Thus
for this choice of parametrization we can increase the expected number of
edges from~$i$ to all other vertices by increasing the magnitude~$k_i$ of
the vector~$\vec{k}_i$, hence increasing the vertex's degree.  At the same
time we can independently control the relative probability of connections
within groups (when $r=s$) and between them ($r\ne s$) by varying the
angle~$\phi$.

If we set~$\phi=0$ (so that all $\vec{v}_r$ point in the $(1,1,1,\ldots)$
direction) then this model becomes equivalent to the variant of the
configuration model in which the expected vertex degrees are fixed and
there is probability~$k_ik_j/2m$ of connection between each pair of
vertices, regardless of community membership.  (Alternatively, if we set
the number of groups $q$ to~1, so that the vectors~$\vec{k}_i$ become
scalars~$k_i$ then we also recover the configuration model.)  If we allow
$\phi$ to be nonzero but make all $k_i$ equal to the same constant
value~$a$, then the model becomes equivalent to the standard stochastic
block model, having a probability $\pin = a^2/2m$ of connection between
vertices in the same community and a smaller probability $\pout = (a^2/2m)
\cos\phi$ between vertices in different communities.  For all other
choices, the model generalizes both the configuration model and the
stochastic block model, allowing us to have nontrivial degrees and
community structure in the same network, as well as other more complex
types of structure (such as overlapping groups---see Ref.~\cite{BKN11}).

\section{Calculation of the spectrum}
In this section we calculate the average spectrum of the adjacency
matrix~$\mat{A}$ for networks generated from the model above, in the limit
of large system size.  The adjacency matrix is the symmetric matrix with
elements~$A_{ij}$ equal to the number of edges between vertices~$i$
and~$j$.  The elements are Poisson independent random integers for our
model, although crucially they are not identically distributed.  The
spectra of matrices with Poisson elements of this kind can be calculated
using methods of random matrix theory.  Our strategy will be first to
calculate the spectrum of the matrix
\begin{equation}
\mat{X} = \mat{A} - \av{\mat{A}},
\label{eq:centered}
\end{equation}
where $\av{\mat{A}}$ is the average value of the adjacency matrix within
the model, which has elements $\av{A_{ij}} = \vec{k}_i\ccdot\vec{k}_j/2m$.
Since $\vec{k}_i$ is a $q$-element vector, this implies that~$\av{\mat{A}}$
has rank~$q$ and hence its eigenvector decomposition has the form
\begin{equation}
\av{\mat{A}} = \sum_{r=1}^q \alpha_r \vec{u}_r\vec{u}_r^T,
\label{eq:ava}
\end{equation}
where $\vec{u}$ are normalized eigenvectors and~$\alpha_r$ are the
corresponding eigenvalues.

The matrix~$\mat{X}$ is a ``centered'' random matrix, having independent
random elements with zero mean, which makes the calculation of its spectrum
particularly straightforward.  Once we have calculated the spectrum of this
centered matrix we will then add the rank-$q$ term~$\av{\mat{A}}$ back in
as a perturbation:
\begin{equation}
\mat{A} = \mat{X} + \av{\mat{A}}.
\label{eq:perturb}
\end{equation}
As we will see, the only property of the centered matrix needed to compute
its spectrum is the variance of its elements, and since the variance of a
Poisson distribution is equal to its mean, we can immediately deduce that
the variance of the $ij$ element of~$\mat{X}$ is
$\vec{k}_i\ccdot\vec{k}_j/2m$.

\subsection{Spectrum of the centered matrix}
In this section we calculate the spectral density~$\rho(z)$ of the centered
matrix~$\mat{X}$, Eq.~\eqref{eq:centered}.  The spectral density is defined
by
\begin{equation}
\rho(z) = {1\over n} \sum_{i=1}^n \delta(z-\lambda_i),
\end{equation}
where $\lambda_i$ is the $i$th eigenvalue of $\mat{X}$ and $\delta(z)$ is
the Dirac delta.  The starting point for our calculation is the well-known
Stieltjes--Perron formula, which gives the spectral density directly in
terms of the matrix as
\begin{equation}
\rho(z) = -{1\over n\pi} \Im \tr\bigl\langle (z-\mat{X})^{-1}
                             \bigr\rangle,
\label{eq:spformula}
\end{equation}
where $z-\mat{X}$ is shorthand for $z\mat{I}-\mat{X}$ with $\mat{I}$ being
the identity.

To calculate the trace, we follow the approach of Bai and
Silverstein~\cite{BS10}, making use of the result that the~$i$th diagonal
component of the inverse of a symmetric matrix~$\mat{B}$ is~\cite{NN13}
\begin{equation}
\bigl[ \mat{B}^{-1} \bigr]_{ii} = {1\over B_{ii} -
                                  \vec{b}_i^T\mat{B}_i^{-1}\vec{b}_i},
\label{eq:tao1}
\end{equation}
where $B_{ii}$ is the $i$th diagonal element of~$\mat{B}$, $\vec{b}_i$~is
the $i$th column of the matrix, and $\mat{B}_i$~is the matrix with the
$i$th row and column removed.  In the limit of large system size, and
provided that the degrees of vertices become large as the network does, the
distribution of values of $[ \mat{B}^{-1} ]_{ii}$ becomes narrowly peaked
about its mean, and one can write the mean value as
\begin{equation}
\bigl\langle \bigl[ \mat{B}^{-1} \bigr]_{ii} \bigr\rangle
 = {1\over\av{B_{ii}} - \av{\vec{b}_i^T\mat{B}_i^{-1}\vec{b}_i}}.
\label{eq:tao2}
\end{equation}

If, as in our case, the elements of~$\mat{B}$ are independent random
variables with mean zero, then
\begin{align}
\bigl\langle \vec{b}_i^T\mat{B}_i^{-1}\vec{b}_i \bigr\rangle
  &= \sum_{jk} \bigl\langle \bigl[ \mat{B}_i^{-1} \bigr]_{jk} \bigr\rangle
     \bigl\langle [\vec{b}_i]_j [\vec{b}_i]_k \bigr\rangle \nonumber\\
  &= \sum_j \bigl\langle \bigl[ \mat{B}_i^{-1} \bigr]_{jj} \bigr\rangle
     \bigl\langle [\vec{b}_i]_j^2 \bigr\rangle,
\end{align}
where we have made use of $\av{[\vec{b}_i]_j [\vec{b}_i]_k} =
\av{[\vec{b}_i]_j} \av{[\vec{b}_i]_k} = 0$ when $j\ne k$.

In our particular example we have $\mat{B}=z-\mat{X}$, which means that
\begin{equation}
[\vec{b}_i]_j = - X_{ij}
\end{equation}
(since $i\ne j$ by definition, the $i$th row having been removed from the
matrix), so
\begin{align}
\bigl\langle \vec{b}_i^T\mat{B}_i^{-1}\vec{b}_i \bigr\rangle
  &= \sum_j \bigl\langle \bigl[ \mat{B}_i^{-1} \bigr]_{jj} \bigr\rangle
     \bigl\langle X_{ij}^2 \bigr\rangle
   = \sum_j
     \bigl\langle \bigl[ \mat{B}_i^{-1} \bigr]_{jj} \bigr\rangle\>
     {\vec{k}_i\ccdot\vec{k}_j\over2m} \nonumber\\
  &= {1\over2m} \vec{k}_i\ccdot \sum_j \vec{k}_j
     \bigl\langle \bigl[ (z-\mat{X})^{-1} \bigr]_{jj} \bigr\rangle,
\end{align}
where the last equality applies in the limit of large system size (for
which it makes a vanishing difference whether we drop the $i$th row and
column from the matrix or not, so $\mat{B}_i$ can be replaced with
$z-\mat{X}$ for all~$i$).  Then, noting that $\av{B_{ii}} = z - \av{X_{ii}}
= z$, Eq.~\eqref{eq:tao2} becomes
\begin{equation}
\bigl\langle \bigl[ (z-\mat{X})^{-1} \bigr]_{ii} \bigr\rangle
  = {1\over z - \vec{k}_i\ccdot \sum_j \vec{k}_j
     \bigl\langle \bigl[ (z-\mat{X})^{-1} \bigr]_{jj} \bigr\rangle/2m}.
\label{eq:tao3}
\end{equation}
Summing this expression over~$i$ we then get the trace we were looking for,
which we will write in terms of a new function
\begin{align}
g(z) &= {1\over n} \tr \bigl\langle (z-\mat{X})^{-1} \bigr\rangle
   = {1\over n} \sum_{i=1}^n
     \bigl\langle \bigl[ (z-\mat{X})^{-1} \bigr]_{ii} \bigr\rangle \nonumber\\
  &= {1\over n} \sum_{i=1}^n {1\over z - \vec{k}_i\ccdot\vec{h}(z)},
\label{eq:stieltjes1}
\end{align}
where we have for convenience defined the vector function
\begin{equation}
\vec{h}(z) = {1\over2m} \sum_i \vec{k}_i
     \bigl\langle \bigl[ (z-\mat{X})^{-1} \bigr]_{ii} \bigr\rangle.
\label{eq:hz1}
\end{equation}
The quantity $g(z)$ (which is just the trace divided by~$n$) is called the
\defn{Stieltjes transform} of the matrix~$\mat{X}$, and it will play a
substantial role in the remainder of our calculation.

It remains to calculate the function~$\vec{h}(z)$, which is now
straightforward.  Multiplying Eq.~\eqref{eq:tao3} by~$\vec{k}_i$ and
substituting into~\eqref{eq:hz1}, we get
\begin{equation}
\vec{h}(z) = {1\over2m} \sum_i {\vec{k}_i\over z - \vec{k}_i\ccdot\vec{h}(z)}.
\label{eq:hz2}
\end{equation}
The solution for the spectral density involves solving this equation
for~$\vec{h}(z)$, then substituting the answer into
Eq.~\eqref{eq:stieltjes1} to get the Stieltjes transform~$g(z)$.  Then the
spectral density itself can be calculated from Eq.~\eqref{eq:spformula}:
\begin{equation}
\rho(z) = - {1\over\pi} \Im g(z).
\label{eq:rhoz1}
\end{equation}

Alternatively, we can simplify the calculation somewhat by rewriting
Eq.~\eqref{eq:tao3} as
\begin{equation}
z\bigl\langle \bigl[ (z-\mat{X})^{-1} \bigr]_{ii} \bigr\rangle
  - \bigl\langle \bigl[ (z-\mat{X})^{-1} \bigr]_{ii} \bigr\rangle \,
    \vec{k}_i\ccdot\vec{h}(z) = 1,
\end{equation}
then summing over~$i$ and dividing by~$n$ to get
$z g(z) - c \|\vec{h}(z)\|^2 = 1$, or
\begin{equation}
g(z) = {1 + c\|\vec{h}(z)\|^2\over z},
\label{eq:gzhz}
\end{equation}
where $c=2m/n$ as previously, which is the average degree of the network,
and $\|\vec{h}(z)\|$ denotes the vector magnitude of~$\vec{h}(z)$,
i.e.,~$\vec{h}\cdot\vec{h}$ (not the complex absolute value).  Then the
spectral density itself, from Eq.~\eqref{eq:rhoz1}, is
\begin{equation}
\rho(z) = - {c\over\pi z} \Im \|\vec{h}(z)\|^2.
\label{eq:rhoz2}
\end{equation}

If we further suppose that the parameter vectors~$\vec{k}_i$ are drawn
independently from some probability distribution~$p(\vec{k})$, which plays
roughly the role played by the degree distribution in other network models,
then in the limit of large network size Eq.~\eqref{eq:hz2} can be written
as
\begin{equation}
\vec{h}(z) = {1\over c} \int {\vec{k}\,p(\vec{k}) \>\dd^q k\over
             z - \vec{k}\ccdot\vec{h}(z)}.
\label{eq:hz3}
\end{equation}

Equations~\eqref{eq:rhoz2} and~\eqref{eq:hz3} between them give us our
solution for the spectral density.  These equations can be regarded as
generalizations of the equations for the configuration model given in
Ref.~\cite{NN13} and similar equations have also appeared in applications
of random matrix methods to other
problems~\cite{molchanov1992limiting,shlyakhtenko1996random,anderson2006clt,casati2009wigner,bai2010limiting}.

\subsection{Examples}
\label{sec:examples1}
As an example of the methods of the previous section, consider a network of
$n$ vertices with two communities of $\half n$ vertices each.  Let the
first group consist of vertices $1\ldots\half n$ and the second of vertices
$\half n+1\ldots n$.  Vertices in the first group will have parameter
vector~$\vec{k}_i = (\kappa_i,\theta)$ and those in the second group will
have $\vec{k}_i = (\kappa_{i-n/2},-\theta)$, where the
quantities~$\kappa_i$ and~$\theta$ are positive constants that we choose
and $\kappa_i\ge\theta$ for all~$i$, to ensure that the expected values
$\av{A_{ij}}=\vec{k}_i\ccdot\vec{k}_j/2m$ of the adjacency matrix elements
are non-negative.

This particular parametrization is attractive for a number of reasons.
First, it already takes the form of the rank-2 eigenvector decomposition of
Eq.~\eqref{eq:ava}, which simplifies the our calculations---the two
(unnormalized) eigenvectors are the $n$-element vectors
$(\boldsymbol{\kappa},\boldsymbol{\kappa})$ and $(1,1,\ldots,-1,-1,\dots)$
where $\boldsymbol{\kappa}$ is the $(\half n)$-element vector with elements
$\kappa_1,\ldots,\kappa_{n/2}$.  Also the expected degrees take a
particularly simple form.  The expected degree of vertex~$i$ for $i\le\half
n$ is
\begin{align}
{1\over2m} \sum_{j=1}^n \vec{k}_i\ccdot\vec{k}_j
  &= {1\over2m} \Biggl[ \sum_{j=1}^{n/2} (\kappa_i\kappa_j + \theta^2)
     \nonumber\\
  &+ \sum_{j=n/2+1}^n (\kappa_i\kappa_{j-n/2} - \theta^2) \Biggr]
  = {\kappa_i\over m} \sum_{j=1}^{n/2} \kappa_j.
\end{align}
But, applying Eq.~\eqref{eq:twom}, we have $m = \sum_{j=1}^{n/2} \kappa_j$
and hence the expected degree of vertex~$i$ is simply~$\kappa_i$.  By a
similar calculation it can easily be shown that for $i>\half n$ the
expected degree is $\kappa_{i-n/2}$, and the average degree in the whole
network is
\begin{equation}
c = {1\over n/2} \sum_{i=1}^{n/2} \kappa_i.
\end{equation}
The parameter $\theta$ also has a simple interpretation in this model: it
controls the strength of the community structure.  For instance, when
$\theta=0$ vertices in the two communities are equivalent and there is no
community structure at all.

To calculate the spectrum for this model, we substitute the values
of~$\vec{k}_i$ into Eq.~\eqref{eq:hz3} to get equations for the two
components of the vector function~$\vec{h}(z)$ thus:
\begin{align}
\label{eq:exh1}
h_1(z) &= {1\over c} \int \kappa p(\kappa) \biggl[
          {1\over z - \kappa h_1(z) - \theta h_2(z)} \nonumber\\
       &\hspace{7em}{} + {1\over z - \kappa h_1(z) + \theta h_2(z)} \biggr]
                    \>\dd\kappa, \\
\label{eq:exh2}
h_2(z) &= {\theta\over c} \int p(\kappa) \biggl[
          {1\over z - \kappa h_1(z) - \theta h_2(z)} \nonumber\\
       &\hspace{7em}{} - {1\over z - \kappa h_1(z) + \theta h_2(z)} \biggr]
                    \>\dd\kappa,
\end{align}
where $p(\kappa)$ is the probability distribution of the
quantities~$\kappa_i$.  Equation~\eqref{eq:exh2} has the trivial solution
$h_2(z)=0$, so the two equations simplify to a single one:
\begin{equation}
h_1(z) = {1\over c} \int {\kappa p(\kappa) \>\dd\kappa
                          \over z - \kappa h_1(z)},
\label{eq:h1soln}
\end{equation}
and then
\begin{equation}
\rho(z) = -{c\over\pi z} \Im h_1^2(z),
\label{eq:modelrho}
\end{equation}
which is independent of the parameter~$\theta$.  These results are
identical to those for the corresponding quantities in the ordinary
configuration model with no community structure and expected degree
distribution~$p(\kappa)$, as derived in Ref.~\cite{NN13}, and hence we
expect the spectrum of the centered adjacency matrix to be the same for the
current model as it is for the configuration model with the same
distribution of expected degrees.

\begin{figure*}
\begin{center}
\includegraphics[width=\textwidth]{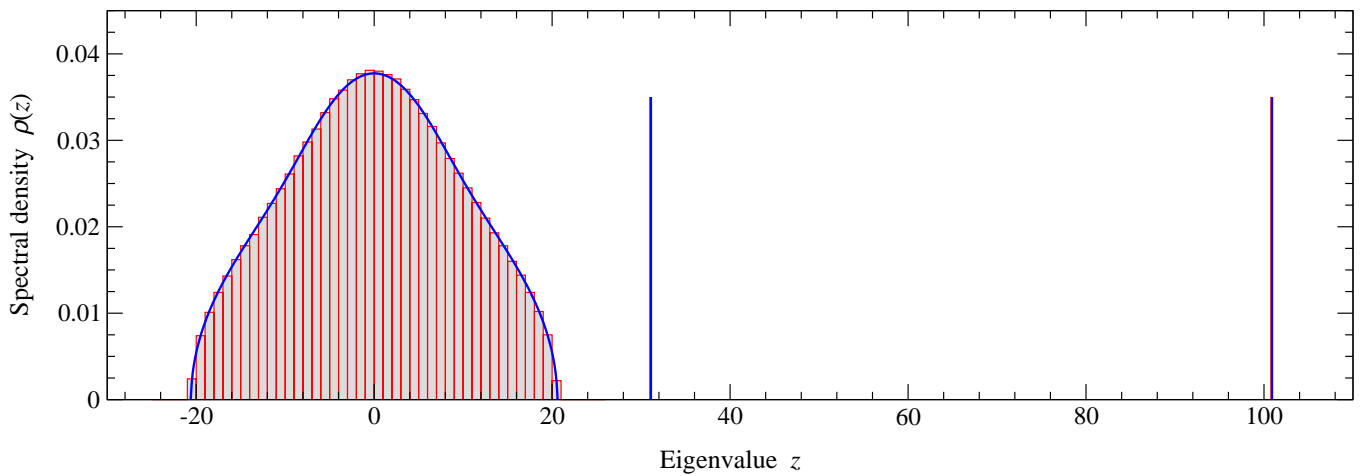}
\end{center}
\caption{The spectrum of the adjacency matrix for the case of a network
  with two groups of equal size and~$\vec{k}_i=(\kappa_i,\pm\theta)$, where
  $\theta=50$, $\kappa_{i+n/2}=\kappa_i$, and $\kappa_i$ is either 60 or
  120 with equal probability.  Blue represents the analytic solution,
  Eqs.~\eqref{eq:sbmhz} and~\eqref{eq:outliers}.  Red is the numerical
  diagonalization of the adjacency matrix of a single network with
  $n=10\,000$ vertices generated from the model with the same parameters.
  The numerically evaluated positions of the two outlying eigenvalues (the
  red spikes) agree so well with the analytic values (blue spikes) that the
  red is mostly obscured behind the blue.}
\label{fig:example}
\end{figure*}

To give a simple example application, suppose that there are only two
different values of~$\kappa$.  Half the vertices in each community have a
value~$\kappa_1$ and the other half~$\kappa_2$.  Then $p(\kappa) =
\half[\delta(\kappa-\kappa_1) + \delta(\kappa-\kappa_2)]$, where
$\delta(x)$ is the Dirac delta, and $c = \half(\kappa_1+\kappa_2)$.  With
this choice
\begin{equation}
h_1(z) = {1\over\kappa_1+\kappa_2} \biggl[ {\kappa_1\over z-\kappa_1 h_1(z)}
         + {\kappa_2\over z-\kappa_2 h_1(z)} \biggr],
\label{eq:sbmhz}
\end{equation}
which can be rearranged to give the cubic equation:
\begin{equation}
\kappa_1\kappa_2 h_1^3 - (\kappa_1+\kappa_2) z h_1^2
  + \biggl[ {2\kappa_1\kappa_2\over\kappa_1+\kappa_2} + z^2 \biggr] h_1
  - z = 0,
\label{eq:h1cubic}
\end{equation}
which can be solved exactly for $h_1(z)$ and hence we can derive an exact
expression for the spectral density.  The expression itself is cumbersome
(like the solutions of most cubic equations), but Fig.~\ref{fig:example}
shows an example for the choice $\kappa_1=60$, $\kappa_2=120$, along with
numerical results for the spectrum of a single random realization of the
model.  As the figure shows, the two agree well.  (The histogram in the
left-hand part of the figure represents the spectrum of the centered
matrix.  The two outlying eigenvalues that appear to the right belong to
the full, non-centered adjacency matrix and are calculated in the following
section.)

Note also that in the special case where $\kappa_1=\kappa_2=c$, so
that~$\kappa$ is constant over all vertices, Eq.~\eqref{eq:sbmhz}
simplifies further to
\begin{equation}
h_1(z) = {1\over z-c h_1(z)},
\end{equation}
which is a quadratic equation with solutions
\begin{equation}
h_1(z) = {z \pm \sqrt{z^2-4c}\over2c},
\label{eq:simplehz}
\end{equation}
and hence the spectral density is
\begin{equation}
\rho(z) = {\sqrt{4c-z^2}\over2\pi c},
\label{eq:semicircle}
\end{equation}
where we take the negative square root in Eq.~\eqref{eq:simplehz} to get a
positive density.  Equation~\eqref{eq:semicircle} has the form of the
classic semicircle distribution for random matrices.  This model is
equivalent to the standard stochastic block model and~\eqref{eq:semicircle}
agrees with the expression for the spectral density derived for that model
by other means in Ref.~\cite{NN12}.

\subsection{Spectrum of the adjacency matrix}
So far we have derived the spectral density of the centered adjacency
matrix $\mat{X} = \mat{A} - \av{\mat{A}}$.  We can use the results of these
calculations to compute the spectrum of the full adjacency matrix by
generalizing the method used in~\cite{NN13}, as follows.

Using Eq.~\eqref{eq:ava} we can write the adjacency matrix as
\begin{equation}
\mat{A} = \mat{X} + \av{\mat{A}} = \mat{X} + 
  \sum_{r=1}^q \alpha_r \vec{u}_r\vec{u}_r^T.
\label{eq:fulla}
\end{equation}
Let us first consider the effect of adding just one of the terms in the sum
to the centered matrix~$\mat{X}$, calculating the spectrum of the matrix
$\mat{X} + \alpha_1 \vec{u}_1\vec{u}_1^T$.  Let $\vec{v}$ be an eigenvector
of this matrix with eigenvalue~$z$:
\begin{equation}
(\mat{X} + \alpha_1 \vec{u}_1\vec{u}_1^T)\vec{v} = z\vec{v}.
\end{equation}
Rearranging this equation we have $\alpha_1 \vec{u}_1\vec{u}_1^T\vec{v} =
(z-\mat{X})\vec{v}$ and, multiplying by $\vec{u}_1^T(z-\mat{X})^{-1}$, we
find
\begin{equation}
\vec{u}_1^T(z-\mat{X})^{-1}\vec{u}_1 = {1\over\alpha_1}.
\end{equation}
Note that the vector~$\vec{v}$ has canceled out of the equation, leaving
us with an equation in~$z$ alone.  The solutions for~$z$ of this equation
give us the eigenvalues of the matrix $\mat{X} + \alpha_1
\vec{u}_1\vec{u}_1^T$.

\begin{figure}[b]
\begin{center}
\includegraphics[width=\columnwidth]{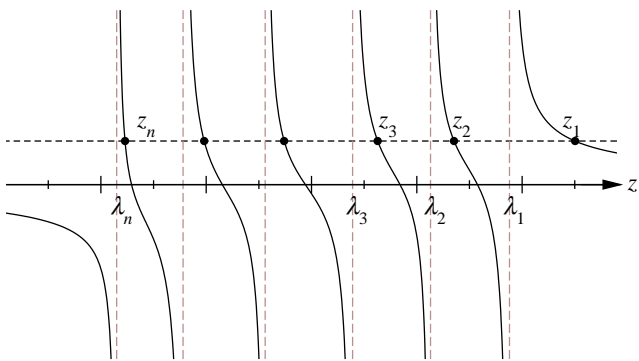}
\end{center}
\caption{A plot of the left-hand size of Eq.~\eqref{eq:interlace1} as a
  function of~$z$ has simple poles at $z=\lambda_i$ for all~$i$.  The
  solutions of the equation fall at the points where the curve crosses the
  horizontal dashed line representing the value of~$1/\alpha_1$.  From the
  geometry of the figure we can see that the solutions must lie in between
  the values of the~$\lambda_i$, interlacing with them, so that
  $z_1\ge\lambda_1\ge z_2\ge\ldots\ge z_n\ge\lambda_n$.}
\label{fig:solution}
\end{figure}

Expanding the vector~$\vec{u}_1$ as a linear combination of the
eigenvectors~$\vec{x}_i$ of the matrix~$\mat{X}$, the equation can also be
written in the form
\begin{equation}
\sum_{i=1}^n {(\vec{x}_i^T\vec{u}_1)^2\over z-\lambda_i} = {1\over\alpha_1},
\label{eq:interlace1}
\end{equation}
where $\lambda_i$ are the eigenvalues of~$\mat{X}$.
Figure~\ref{fig:solution} shows a graphical representation of the solution
of this equation for the eigenvalues~$z$.  The left-hand side of the
equation, represented by the solid curves, has simple poles at
$z=\lambda_i$ for all~$i$.  The right-hand side, represented by the
horizontal dashed line, is constant.  Where the two intercept, represented
by the dots, are the solutions for~$z$.  From the geometry of the figure we
can see that the values of~$z$ must fall between consecutive values
of~$\lambda_i$---we say that the $z$'s and $\lambda$'s are
\defn{interlaced}.  If we number the eigenvalues~$\lambda_i$ in order from
largest to smallest so that $\lambda_1\ge\lambda_2\ge\ldots\ge\lambda_n$,
and similarly for the $n$ solutions~$z_i$ to Eq.~\eqref{eq:interlace1},
then $z_1\ge\lambda_1\ge z_2\ge\lambda_2\ge\ldots\ge z_n\ge\lambda_n$.  In
the limit of large system size, as the~$\lambda_i$ become more and more
closely spaced in the spectrum of the matrix, this interlacing places
tighter and tighter bounds on the values of~$z_i$, and asymptotically we
have $z_i=\lambda_i$ and the spectral density of $\mat{X} + \alpha_1
\vec{u}_1\vec{u}_1^T$ is the same as that of $\mat{X}$ alone.

There is one exception, however, in the highest-lying eigenvalue~$z_1$,
which is bounded below by~$\lambda_1$ but unbounded above, meaning it need
not be equal to~$\lambda_1$ and may lie outside the band of values occupied
by the spectrum of the matrix~$\mat{X}$.  To calculate this eigenvalue we
observe that, the matrix~$\mat{X}$ being random, its
eigenvectors~$\vec{x}_i$ are also random and hence $\vec{x}_i^T\vec{u}_1$
is a zero-mean random variable with variance~$1/n$.  Taking the average of
Eq.~\eqref{eq:interlace1} over the ensemble of networks, the numerator on
the left-hand side gives simply a factor of $1/n$ and we have
\begin{equation}
{1\over\alpha_1}
  = {1\over n} \Biggl\langle\sum_{i=1}^n {1\over z-\lambda_i} \Biggr\rangle
  = {1\over n} \bigl\langle \tr(z-\mat{X})^{-1} \bigr\rangle
  = g(z).
\label{eq:z1}
\end{equation}
The solution to this equation gives us the value of~$z_1$.

This then gives us the complete spectrum for the matrix $\mat{X} + \alpha_1
\vec{u}_1\vec{u}_1^T$.  It consists of a continuous spectral band with
spectral density equal to that of the matrix~$\mat{X}$ alone, which is
calculated from Eq.~\eqref{eq:rhoz2}, plus a single eigenvalue outside the
band whose value is the solution for~$z$ of~$g(z)=1/\alpha_1$.

\begin{figure}
\begin{center}
\includegraphics[width=8cm]{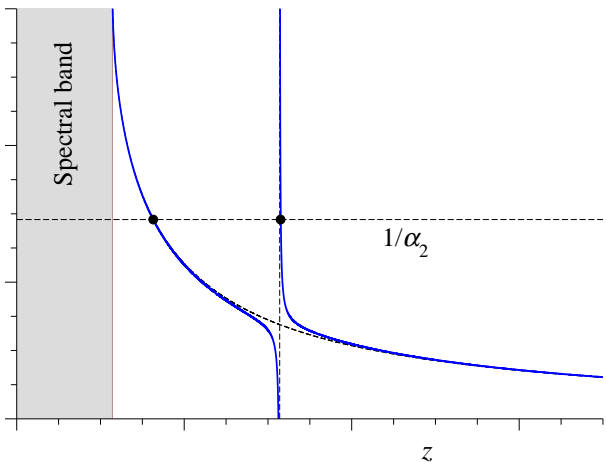}
\end{center}
\caption{A graphical representation of the solution of
  Eq.~\eqref{eq:zprime}.  The left-hand size of the equation, represented
  by the solid blue curve, follows closely the form of the Stieltjes
  transform~$g(z)$, except within a distance of order~$1/n$ from~$z_1$,
  where it diverges.  The horizontal dashed line represents the
  value~$1/\alpha_2$ and the solutions to~\eqref{eq:zprime}, of which there
  are two, fall at the intersection of this line with the solid curve, as
  indicated by the dots.  One of these solutions coincides closely
  with~$z_1$, the other is the solution of $g(z)=1/\alpha_2$.}
\label{fig:outliers}
\end{figure}

We could have made the same argument about any single term
$\alpha_r\vec{u}_r\vec{u}_r^T$ appearing in Eq.~\eqref{eq:fulla} and
derived the corresponding result that the continuous spectral band is
unchanged from the centered matrix but there can be an outlying eigenvalue
$z_r$ given by
\begin{equation}
g(z_r) = {1\over\alpha_r}.
\label{eq:outliers}
\end{equation}
The calculation of the spectrum of the full adjacency matrix requires that
we consider all terms in Eq.~\eqref{eq:fulla} simultaneously, but in
practice it turns out that it is enough to consider them one by one using
Eq.~\eqref{eq:outliers}.  The argument for this is in two parts as follows.
\begin{enumerate}
\item We have shown that the spectral density of the continuous band in the
  spectrum of the matrix $\mat{X} + \alpha_1 \vec{u}_1\vec{u}_1^T$ is the
  same as that for the matrix~$\mat{X}$ alone, and there is one additional
  outlying eigenvalue, which we denote~$z_1$.  Now we can add another term
  $\alpha_2\vec{u}_2\vec{u}_2^T$ and repeat our argument for the matrix
  $\mat{X} + \alpha_1 \vec{u}_1\vec{u}_1^T + \alpha_2\vec{u}_2\vec{u}_2^T$,
  finding the equivalent of Eq.~\eqref{eq:interlace1} to be
\begin{equation}
\sum_{i=2}^n {(\vec{x}_i^T\vec{u}_2)^2\over z'-z_i}
    + {(\vec{x}_1^T\vec{u}_2)^2\over z'-z_1}
  = {1\over\alpha_2},
\label{eq:interlace2}
\end{equation}
where $z'$ is the eigenvalue of the new matrix and $z_i$ are the solutions
of~\eqref{eq:interlace1}.  As before, this implies there is an interlacing
condition and that the spectral density of the perturbed matrix is the same
within the spectral band as that for the unperturbed matrix.  We can repeat
this argument as often as we like and thus demonstrate that the shape of
the spectral band never changes, so long as the number of perturbations
(which is also the rank of~$\av{\mat{A}}$) is small compared to the size of
the network, i.e,~$q\ll n$.
\item This argument pins down all but the top two eigenvalues of $\mat{X} +
  \alpha_1 \vec{u}_1\vec{u}_1^T + \alpha_2\vec{u}_2\vec{u}_2^T$.  These two
  we can calculate by a variant of our previous argument.  We average
  Eq.~\eqref{eq:interlace2} over the ensemble, noting again that
  $\av{(\vec{x}_i^T\vec{u}_2)^2} = 1/n$ and find that
\begin{equation}
{1\over n} \sum_{i=2}^n {1\over z'-z_i}
    + {1/n\over z'-z_1} = {1\over\alpha_2}.
\end{equation}
For large~$n$ the first sum is once again equal to the Stieltjes
transform~$g(z)$ and hence the top two eigenvalues are solutions for~$z'$
of
\begin{equation}
g(z') + {1/n\over z'-z_1} = {1\over\alpha_2}.
\label{eq:zprime}
\end{equation}
But $g(z)$ and $\alpha_2$ are of order~1, while the term $n^{-1}/(z'-z_1)$
is of order~$1/n$ and hence can in most circumstances be neglected, giving
$g(z') = 1/\alpha_2$, which recovers Eq.~\eqref{eq:outliers}.  The only
time this term cannot be neglected is when $z'$ is within a distance of
order~$1/n$ from~$z_1$, in which case we have a simple pole in the
left-hand side of the equation as $z'$ approaches~$z_1$.  Thus the
left-hand side has the form sketched in Fig.~\ref{fig:outliers}, following
$g(z)$ closely for most values of~$z$, but diverging suddenly when very
close to~$z_1$.  Equation~\eqref{eq:zprime} then has two solutions, as
indicated by the dots in the figure, one given by $g(z)=1/\alpha_2$ and one
that is asymptotically equal to~$z_1$, which is the solution of
$g(z)=1/\alpha_1$.
\end{enumerate}

We can repeat this argument as many times as we like to demonstrate that
the outlying eigenvalues are just the $q$ solutions of
Eq.~\eqref{eq:outliers} for each value~$r=1\ldots q$.  Thus our final
solution for the complete spectrum of the adjacency matrix has two parts: a
continuous spectral band, given by Eqs.~\eqref{eq:rhoz2}
and~\eqref{eq:hz3}, and $q$ out\-lying eigenvalues, given by the solutions
of Eq.~\eqref{eq:outliers}, with $g(z)$ given by Eq.~\eqref{eq:gzhz}.

\subsection{Examples}
\label{sec:examples2}
Let us return to the examples of Section~\ref{sec:examples1} and apply the
methods above to the calculation of their outlying eigenvalues.  Recall
that we looked at networks with two communities and chose parameter
vectors~$\vec{k}_i = (\kappa_i,\theta)$ for vertices in the first community
and $\vec{k}_i = (\kappa_{i-n/2},-\theta)$ for those in the second.  For
such networks the vector function~$\vec{h}(z)$ reduces to a single scalar
function~$h_1(z)$ that satisfies Eq.~\eqref{eq:h1soln}.  At the same time,
Eq.~\eqref{eq:gzhz} tells us that for this model $zg(z) = 1+c h_1^2(z)$ and
hence from Eq.~\eqref{eq:outliers} the positions of the outlying
eigenvalues are solutions of
\begin{equation}
1 + c h_1^2(z) - {z\over\alpha_r} = 0,
\label{eq:houtliers}
\end{equation}
for $r=2\ldots q$.  Locating the outliers is thus a matter of solving
\eqref{eq:h1soln} for~$h_1$, substituting the result
into~\eqref{eq:houtliers}, and then solving for~$z$.

Consider, for instance, the choice we made in Section~\ref{sec:examples1},
where there were just two values of~$\kappa$, denoted $\kappa_1$
and~$\kappa_2$, with half the vertices in each community taking each value.
Then $h_1$ obeys the cubic equation~\eqref{eq:h1cubic}, which can be solved
exactly, and hence we can calculate the position of the outliers.
Figure~\ref{fig:example} shows the results for the choice $\kappa_1=60$,
$\kappa_2=120$, $\theta=50$, along with numerical results for the same
parameter values.  As the figure shows, analytic and numerical calculations
again agree well---so well, in fact, that the difference between them is
quite difficult to make out on the plot.

We also looked in Section~\ref{sec:examples1} at the simple case where
$\kappa=c$ for all vertices, so that they all have the same expected
degree, in which case the model becomes equivalent to the standard
stochastic block model and the continuous spectral band takes the classic
semicircle form of Eq.~\eqref{eq:semicircle}.  For this model we have
$\alpha_1=c$ and $\alpha_2=\theta^2/c$.  Using Eq.~\eqref{eq:simplehz} for
$h_1(z)$ and solving~\eqref{eq:houtliers} for~$z$, we then find the top two
eigenvalues of the adjacency matrix to be
\begin{equation}
z_1 = c + 1, \qquad
z_2 = {\theta^2\over c} + {c^2\over\theta^2},
\end{equation}
which agrees with the results given previously for the stochastic block
model in Ref.~\cite{NN12}.

\subsection{Detectability of communities}
One of the primary uses of network spectra is for the detection of
community structure~\cite{Newman06c,NN12}.  As we have seen, the number of
eigenvalues above the edge of the spectral band is equal to the number of
communities in the network, and hence the observation of these eigenvalues
can be taken as evidence of the presence of communities and their number as
an empirical measure of the number of communities.  The identity of the
communities themselves---which vertices belong to which community---can be
deduced, at least approximately, by looking at the elements of the
eigenvectors~\cite{Newman06c}.

However, as shown previously in~\cite{NN12} for the simplest two-community
block model, the position of the leading eigenvalues varies as one varies
the strength of community structure, and for sufficiently low (but still
nonzero) strength an eigenvalue may meet the edge of the spectral band and
hence become invisible in the spectrum, meaning it can no longer be used as
evidence of the presence of community structure.  Moreover, as also shown
in~\cite{NN12}, the elements of the corresponding eigenvector become
uncorrelated with group membership at this point, so that any algorithm
which identifies communities by examining the eigenvector elements will
fail.  The point where this happens, at least in the simple two-community
model, coincides with the known ``detectability threshold'' for community
structure, at which it is believed all algorithms for community detection
must fail~\cite{RL08,DKMZ11a,HRN12}.

We expect qualitatively similar behavior in the present model as well.
Consider the Stieltjes transform~$g(z)$ defined in
Eq.~\eqref{eq:stieltjes1}.  Inside the spectral band the transform is
complex by definition---see from Eq.~\eqref{eq:rhoz1}.  Above the band it
is real and monotonically decreasing in~$z$, as we can see by evaluating
the trace in the basis in which~$\mat{X}$ is diagonal:
\begin{equation}
g(z) = {1\over n} \sum_{i=1}^n {1\over z-\lambda_i},
\end{equation}
where $\lambda_i$ are the eigenvalues of~$\mat{X}$ as previously.  Above
the band, where $z>\lambda_i$ for all~$i$, every term in this sum is
monotonically decreasing, and hence so is~$g(z)$.  This implies via
Eq.~\eqref{eq:outliers} that larger values of~$\alpha_r$ give larger
eigenvalues and that the largest real value~$g_\textrm{max}$ of the
Stieltjes transform occurs exactly at the band edge.  Moreover, as shown in
Ref.~\cite{NN13}, the edge of the band is marked generically by a
square-root singularity in the spectral density, which implies
that~$g_\textrm{max}$ is finite---see Fig.~\ref{fig:outliers} for a sketch
of the function.  Thus when we make the community structure in the network
weaker, meaning we decrease the values of the~$\alpha_r$, we also decrease
the outlying eigenvalues of the adjacency matrix and eventually the lowest
of those eigenvalues will meet the edge of the band and disappear at the
point where $1/\alpha_r=g_\textrm{max}$.  If we continue to weaken the
structure, more eigenvalues will disappear, in order---smallest first, then
second smallest, and so forth.

Thus we expect there to be a succession of detectability transitions in the
network, $q-1$ of them in all, where $q$ again is the number of
communities.  At the first of these transitions the $q$th largest
eigenvalue will meet the band edge and disappear, meaning there will only
be $q-1$ outlying eigenvalues left and hence there will be observational
evidence of only $q-1$ communities in the network, even if in fact we know
there to be~$q$.  At the next transition the number will decrease further
to~$q-2$, and so forth.  One thus loses the ability to detect community
structure in stages, one community at a time.  Final evidence of any
structure at all disappears at the point where the second largest
eigenvalue meets the band edge.

Consider, for instance, the example network from
Section~\ref{sec:examples1} again, in which there are two groups with
parameter vectors of the form~$(\kappa_i,\pm\theta)$, where the
parameters~$\kappa_i$ control the expected degrees and $\theta$ controls
the strength of the community structure.  As before, let us study the case
where the $\kappa_i$ take just two different values with equal probability,
so that $h_1$ satisfies the cubic equation~\eqref{eq:h1cubic} (and
$h_2=0$).  Then we can calculate the maximal real value of~$g(z)$ as
follows.

Like $g(z)$, the function~$h_1(z)$ is real outside the continuous spectral
band but complex inside it, as one can see from Eq.~\eqref{eq:modelrho}.
The band edge is thus the point at which the solution of the cubic equation
becomes complex, which is given by the zero of the discriminant of the
cubic.  Take, for example, the case where $\kappa_1=\kappa$ and
$\kappa_2=2\kappa$ for some constant~$\kappa$.  Then, employing the
standard formula, the discriminant of~\eqref{eq:h1cubic} is
\begin{equation}
{\kappa^5\over27} \biggl[ 27 \biggl( {z^2\over\kappa} \biggr)^3
                   - 216 \biggl( {z^2\over\kappa} \biggr)^2
                   + 252 \biggl( {z^2\over\kappa} \biggr)
                   - 512 \biggr].
\end{equation}
This is zero when, and hence the band edge falls at, $z = \sqrt{x\kappa}$,
where $x\simeq7.058$ is the sole real solution of the cubic equation $27
x^3 - 216 x^2 + 252 x - 512 = 0$.  Substituting into
Eq.~\eqref{eq:h1cubic}, we then find that the value of $h_1$ at the band
edge is $y/\sqrt{\kappa}$ where $y=0.723$ is the smallest real solution of
the cubic equation $2y^3-3\sqrt{x}y^2+(x+\tfrac43)y-\sqrt{x} = 0$.  Then,
using Eq.~\eqref{eq:gzhz} and the fact that the average degree is
$c=\frac32\kappa$, the value of $g(z)$ at the band edge is
\begin{equation}
g_\textrm{max} = {2+3y^2\over2\sqrt{x\kappa}}.
\end{equation}

In this case there is only one parameter $\alpha_r$ with $r\ge2$, which is
$\alpha_2=\theta^2/c$.  Hence there is a single threshold at which we lose
the ability to detect communities, falling~at
\begin{equation}
{c\over\theta^2} = {2+3y^2\over2\sqrt{x\kappa}},
\end{equation}
or
\begin{equation}
\theta = \sqrt{3\sqrt{x\kappa^3}\over2+3y^2}
       \simeq 1.494 \kappa^{3/4}.
\end{equation}
If $\theta$ is smaller than this value then spectral methods will fail to
detect the communities in the network.  We have checked this behavior
numerically and find indeed that spectral community detection fails at
approximately this point.

\section{Conclusions}
In this paper we have given a prescription for calculating the spectrum of
the adjacency matrix of an undirected random network containing both
community structure and a nontrivial degree distribution, generated using
the model of Ball~\etal~\cite{BKN11}.  In the limit of large network size
the spectrum consists in general of two parts: (1)~a continuous spectral
band containing the bulk of the eigenvalues and (2)~$q$ outlying
eigenvalues above the spectral band, where $q$ is the number of communities
in the network.  We give expressions for both the shape of the band and the
positions of the outlying eigenvalues that are exact in the limit of a
large network and large vertex degrees, although their evaluation involves
integrals that may not be analytically tractable in practice, in which case
we must resort to numerical evaluation.  We have compared the spectra
calculated using our method with direct numerical diagonalizations and find
the agreement to be excellent.  Based on our results we also argue that
there should be a series of $q-1$ ``detectability transitions'' as the
community structure gets weaker, at which one's ability to detect
communities becomes successively impaired.  The positions of these
transitions correspond to the points at which the outlying eigenvalues meet
the edge of the spectral band and disappear.  With the disappearance of the
second-largest eigenvalue in this manner, all trace of the community
structure vanishes from the spectrum and the network is indistinguishable
from an unstructured random graph.

\begin{acknowledgments}
  This work was funded in part by the National Science Foundation under
  grants CCF--1116115 and DMS--1107796, by the Air Force Office of
  Scientific Research (AFOSR) and the Defense Advanced Research Projects
  Agency (DARPA) under grant FA9550--12--1--0432, and by the Army Research
  Office under MURI grant W911NF--11--1--0391.
\end{acknowledgments}


\begin{thebibliography}{10}
\expandafter\ifx\csname url\endcsname\relax
  \def\url#1{\texttt{#1}}\fi
\expandafter\ifx\csname urlprefix\endcsname\relax\def\urlprefix{URL }\fi

\bibitem{Newman03d}
M.~E.~J. Newman, The structure and function of complex networks. \textit{SIAM
  Review} \textbf{45}, 167--256 (2003).

\bibitem{Boccaletti06}
S.~Boccaletti, V.~Latora, Y.~Moreno, M.~Chavez, and D.-U. Hwang, Complex
  networks: Structure and dynamics. \textit{Physics Reports} \textbf{424},
  175--308 (2006).

\bibitem{Fiedler73}
M.~Fiedler, Algebraic connectivity of graphs. \textit{Czech. Math. J.}
  \textbf{23}, 298--305 (1973).

\bibitem{PSL90}
A.~Pothen, H.~Simon, and K.-P. Liou, Partitioning sparse matrices with
  eigenvectors of graphs. \textit{SIAM J. Matrix Anal. Appl.} \textbf{11},
  430--452 (1990).

\bibitem{Newman06c}
M.~E.~J. Newman, Finding community structure in networks using the eigenvectors
  of matrices. \textit{Phys. Rev. E} \textbf{74}, 036104 (2006).

\bibitem{Bonacich87}
P.~F. Bonacich, Power and centrality: A family of measures. \textit{Am. J.
  Sociol.} \textbf{92}, 1170--1182 (1987).

\bibitem{FDBV01}
I.~J. Farkas, I.~Der\'enyi, A.-L. Barab\'asi, and T.~Vicsek, Spectra of
  ``real-world'' graphs: Beyond the semicircle law. \textit{Phys. Rev. E}
  \textbf{64}, 026704 (2001).

\bibitem{GKK01a}
K.-I. Goh, B.~Kahng, and D.~Kim, Spectra and eigenvectors of scale-free
  networks. \textit{Phys. Rev. E} \textbf{64}, 051903 (2001).

\bibitem{CLV03}
F.~Chung, L.~Lu, and V.~Vu, Spectra of random graphs with given expected
  degrees. \textit{Proc. Natl. Acad. Sci. USA} \textbf{100}, 6313--6318 (2003).

\bibitem{DGMS03}
S.~N. Dorogovtsev, A.~V. Goltsev, J.~F.~F. Mendes, and A.~N. Samukhin, Spectra
  of complex networks. \textit{Phys. Rev. E} \textbf{68}, 046109 (2003).

\bibitem{Kuhn08}
R.~K{\"u}hn, Spectra of sparse random matrices. \textit{J. Phys. A}
  \textbf{41}, 295002 (2008).

\bibitem{CGO09}
S.~Chauhan, M.~Girvan, and E.~Ott, Spectral properties of networks with
  community structure. \textit{Phys. Rev. E} \textbf{80}, 056114 (2009).

\bibitem{NN12}
R.~R. Nadakuditi and M.~E.~J. Newman, Graph spectra and the detectability of
  community structure in networks. \textit{Phys. Rev. Lett.} \textbf{108},
  188701 (2012).

\bibitem{NN13}
R.~R. Nadakuditi and M.~E.~J. Newman, Spectra of random graphs with arbitrary
  expected degrees. \textit{Phys. Rev. E} \textbf{87}, 012803 (2013).

\bibitem{HLL83}
P.~W. Holland, K.~B. Laskey, and S.~Leinhardt, Stochastic blockmodels: Some
  first steps. \textit{Social Networks} \textbf{5}, 109--137 (1983).

\bibitem{CK01}
A.~Condon and R.~M. Karp, Algorithms for graph partitioning on the planted
  partition model. \textit{Random Structures and Algorithms} \textbf{18},
  116--140 (2001).

\bibitem{RL08}
J.~Reichardt and M.~Leone, ({U}n)detectable cluster structure in sparse
  networks. \textit{Phys. Rev. Lett.} \textbf{101}, 078701 (2008).

\bibitem{DKMZ11a}
A.~Decelle, F.~Krzakala, C.~Moore, and L.~Zdeborov\'a, Inference and phase
  transitions in the detection of modules in sparse networks. \textit{Phys.
  Rev. Lett.} \textbf{107}, 065701 (2011).

\bibitem{HRN12}
D.~Hu, P.~Ronhovde, and Z.~Nussinov, Phase transitions in random {P}otts
  systems and the community detection problem: Spin-glass type and dynamic
  perspectives. \textit{Phil. Mag.} \textbf{92}, 406--445 (2012).

\bibitem{MR95}
M.~Molloy and B.~Reed, A critical point for random graphs with a given degree
  sequence. \textit{Random Structures and Algorithms} \textbf{6}, 161--179
  (1995).

\bibitem{NSW01}
M.~E.~J. Newman, S.~H. Strogatz, and D.~J. Watts, Random graphs with arbitrary
  degree distributions and their applications. \textit{Phys. Rev. E}
  \textbf{64}, 026118 (2001).

\bibitem{CL02b}
F.~Chung and L.~Lu, The average distances in random graphs with given expected
  degrees. \textit{Proc. Natl. Acad. Sci. USA} \textbf{99}, 15879--15882
  (2002).

\bibitem{BKN11}
B.~Ball, B.~Karrer, and M.~E.~J. Newman, An efficient and principled method for
  detecting communities in networks. \textit{Phys. Rev. E} \textbf{84}, 036103
  (2011).

\bibitem{BS10}
Z.~Bai and J.~W. Silverstein, \textit{Spectral analysis of large dimensional
  random matrices}. Springer, Berlin, 2nd edition (2010).

\bibitem{molchanov1992limiting}
S.~Molchanov, L.~Pastur, and A.~Khorunzhii, Limiting eigenvalue distribution
  for band random matrices. \textit{Theoretical and Mathematical Physics}
  \textbf{90}, 108--118 (1992).

\bibitem{shlyakhtenko1996random}
D.~Shlyakhtenko, Random {G}aussian band matrices and freeness with
  amalgamation. \textit{International Mathematics Research Notices}
  \textbf{1996}, 1013--1025 (1996).

\bibitem{anderson2006clt}
G.~Anderson and O.~Zeitouni, A {CLT} for a band matrix model.
  \textit{Probability Theory and Related Fields} \textbf{134}, 283--338
  (2006).

\bibitem{casati2009wigner}
G.~Casati and V.~Girko, Wigner's semicircle law for band random matrices.
  \textit{Random Operators and Stochastic Equations} \textbf{1}, 15--22
  (2009).

\bibitem{bai2010limiting}
Z.~Bai and L.~Zhang, The limiting spectral distribution of the product of the
  {W}igner matrix and a nonnegative definite matrix. \textit{Journal of
  Multivariate Analysis} \textbf{101}, 1927--1949 (2010).

\end{thebibliography}
\end{document}